\documentclass[lettersize,journal]{IEEEtran}
\usepackage{amsmath,amsfonts}
\usepackage{algorithm} 
\usepackage{algorithmic}  
\usepackage[algo2e]{algorithm2e} 
\newcommand{\dblprime}{^{\prime\prime}}
\newcommand{\trpprime}{^{\prime\prime\prime}}
\usepackage{array}
\usepackage[caption=false,font=normalsize,labelfont=sf,textfont=sf]{subfig}
\usepackage{textcomp}
\usepackage{stfloats}
\usepackage{url}
\usepackage{verbatim}
\usepackage{graphicx}
\usepackage{cite}
\usepackage[usestackEOL]{stackengine}
\usepackage{amssymb}
\usepackage{color,soul}         
\usepackage{tablefootnote}      
\begin{document}

\title{High-Throughput Rate-Flexible Combinational Decoders for Multi-Kernel Polar Codes}

\author{Hossein Rezaei,~\IEEEmembership{Student Member,~IEEE,}
        Nandana Rajatheva,~\IEEEmembership{Senior Member,~IEEE}, and
        Matti Latva-aho,~\IEEEmembership{Senior Member,~IEEE}}

\IEEEpubid{0000--0000/00\$00.00~\copyright~2023 IEEE}

\maketitle
\begin{abstract}
Polar codes have received growing attention in the past decade and have been selected as the coding scheme for the control channel in the fifth generation (5G) wireless communication systems. However, the conventional polar codes have only been constructed by binary ($2\times 2$) kernel which poses block length limitation to powers of $2$. To attain more flexible block lengths, multi-kernel polar codes are proposed. 
In this paper, a combinational architecture for multi-kernel polar codes with high throughput is proposed based on successive cancellation decoding algorithm. The proposed scheme can decode pure-binary, pure-ternary ($3\times 3$), and binary-ternary mixed polar codes. The decoder's architecture is rate-flexible meaning that a new code rate can be assigned to the decoder at every clock cycle. The proposed architecture is validated by FPGA implementation and the results reveal that a code of size $N=81$ gains the coded throughput of $1664.5$ Mbps. A novel Python-based polar compiler is also proposed to automatically generate the HDL modules for target decoders. A designer can input the target block length and kernel ordering of a polar code, and get the required VHDL files automatically. Based on our simulations, the majority of the required HDL files can be generated in less than $0.4$ seconds.
\end{abstract}
\begin{IEEEkeywords}Polar code, successive-cancellation decoder, multi-kernel, error-correcting codes, hardware implementation, polar compiler.
\end{IEEEkeywords}
\section{Introduction}
\label{sec_intro}
\IEEEPARstart{P}{olar} codes are introduced by Arikan \cite{Arikan} as a family of error-correcting codes with the capability to achieve symmetric channel capacity of the binary-input discrete memoryless channel when the code length approaches infinity. Using a recursive construction, a polar code of size $N=2^n$ can be constructed by the $n$th Kronecker power of binary matrix $T_2=[\begin{smallmatrix} 1&1\\ 1&0 \end{smallmatrix}]$ also known as Arikan's kernel. This construction converts the physical channel into $N$ virtual synthetic channels whose reliability approaches either zero or infinity as the code length grows. 

Arikan in \cite{Arikan} also proved that polar codes can achieve the symmetric channel capacity using successive cancellation (SC) decoding algorithm. 
Thenceforth, researchers have intensively sought to improve polar codes of limited size in terms of decoding latency under SC, complexity, power, and error-correction performance. Successive cancellation list (SCL) \cite{tal2015list} decoding concatenated with CRC \cite{niu2012crc} improves the error-correction performance of polar codes allowing them to compete with other channel coding methods like low-density parity-check (LDPC) codes. This effort makes the foundation for the adaption of polar codes to the 3GPP fifth generation new radio (5G-NR) wireless communication standards \cite{3GPP}.

The majority of current research however has focused on polar codes constructed by Arikan's kernel \cite{Arikan}. Using a $2\times2$ polarization matrix limits the block length of polar codes to powers of $2$ which can not address all demanding code lengths and rates in beyond 5G networks. The rate-matching schemes such as puncturing and shortening \cite{Han2022, bioglio2017} methods have been proposed to cover non-binary block lengths. However, a priori performance and optimality of punctured and shortened codes are hard to evaluate. The polarization phenomenon attained by the Kronecker products of $2$ can also be expanded to other kennels. The ternary kernel ($3\times 3$) in particular has been receiving increased attention due to its low complexity and offering polarization optimality. Using this method, multi-kernel (MK) polar codes \cite{Xia2020, gabry2017, bioglio} offer flexible code lengths by employing kernels with different dimensions. They are characterized by the same computational complexity as Arikan's polar codes and outperform similar punctured and shortened codes in terms of error-correction performance and complexity \cite{gabry2017, Rezaei2022MK}. 

In terms of hardware implementation, multiple decoders have been proposed for decoding Arikan's polar codes. The work in \cite{Coppolino} adapts the architecture of Arikan's polar codes of \cite{Sarkis} to MK codes constructed by Arikan's and ternary kernels.  An architecture for decoding  MK polar codes with reduced latency is proposed in \cite{Rezaei2022MK}. However, two mentioned MK decoders of \cite{Rezaei2022MK} and \cite{Coppolino} use complex memory interfaces for reading/writing data from/to the memory for binary and ternary stages resulting in reduced coded throughput. 

\IEEEpubidadjcol
The motivation of this paper is to develop a flexible architecture to obtain high-throughput MK polar decoders with low power consumption based on the SC algorithm. We address this motivation by using pure combinational architectures. The recursive and feed-forward structure of the SC algorithm facilitates the adaption of pure combinational logic to polar decoders. The SC-based combinational decoders are fully scalable and operate in considerably lower frequencies compared to their sequential counterparts. However, they are able to decode an entire codeword in one long clock cycle which substantially reduces the dynamic power. 
A key characteristic of the proposed architecture is online rate assignment to a given block length meaning that a new code rate can be assigned to the decoder at every clock cycle. 

In \cite{rezaei2022combinational} we proposed the first MK combinational decoder. It has however two limitations. First, it does not support pure-ternary polar codes. Second, it does not provide the required memory to load the next log-likelihood ratio (LLR) frame and frozen bit indicator, and also to offload the estimated codeword. Lacking these memories results in having limited throughput since the input data is not ready when the previous frame is decoded. Along with detailed performance and complexity analysis, these limitations are addressed in this work. 
The proposed architecture supports 55 different block lengths with a maximum block length of $N_{max} = 4096$ constructed by pure-binary, pure-ternary, and binary-ternary mixed kernels. An FPGA implementation and comparison to state-of-the-art MK decoders is conducted in order to validate the architecture.

Finally, we propose a Python-based hardware compiler to automate the process of generating the VHDL files required for the FPGA implementation of the proposed decoders. The motivation is that by changing the block length or kernel ordering, the entire VHDL modules are subjected to change. Using the proposed polar compiler, once the user inputs the block length and kernel ordering, the compiler automatically outputs all necessary VHDL files. In case the user does not enter a kernel order, the compiler automatically assigns the kernel ordering with the highest error-correction performance.

The remainder of this paper is organized as follows. In Section \ref{sec_back}, we present a background on Arikan's and MK polar codes. The code construction method along with the proposed architecture of the MK decoder and complexity analysis are explained in Section \ref{sec_arch}. Section \ref{sec_compiler} discusses the process of automatic generation of VHDL files for target decoders using the proposed polar compiler. The implementation results and comparison to previous works are detailed in section \ref{sec_res}. Finally, a conclusion will be given in section \ref{sec_conc}.
\section{{Preliminaries}}
\label{sec_back}
In this section along with a background on polar codes, we provide the code construction methods of Arikan's and MK polar codes. Then, the SC algorithms for decoding Arikan's and MK codes will be summarised.
\subsection{Arikan's Polar Codes}
$\mathcal{PC}(N,K)$ denotes a polar code of size $N\text{=}2^n$ with $K$ bits of information where the code rate can be calculated as $\mathcal{R} = {K/N}$. Channel polarization phenomenon is proved by Arikan in \cite{Arikan} for binary polar codes and it can be used to transform the physical channel $W$ into $N$ individual virtual channels $W_{i}^{N}$ ($1 \leq i \leq N$). The divided virtual channels feature relative increased or decreased reliabilities and in case $N \rightarrow \infty$, the reliability of each channel approaches either $1$ (perfectly reliable) or $0$ (perfectly unreliable). Either the Bhattacharya parameters \cite{Arikan} or Gaussian approximation  \cite{mori2009performance} can be used to designate individual reliable channels. The set of $K$ most reliable bit positions is called the information set indicated by $\mathcal{I}$ and the remaining $N\text{-}K$ bit locations are called frozen set denoted by $\mathcal{F}$. The bit values in frozen set locations are set to zero.  

A linear transformation can be used to construct polar codes. It is expressed as $x = uG$ where $x$ indicates the encoded stream, $u$ denotes an $N$-bit input vector to the encoder, and $G$ is the generator matrix. The input vector $u$ is constructed by inserting the message and frozen data into reliable and unreliable positions, respectively. The generator matrix $G = T_2^{\otimes n}$ is constructed by the n-th Kronecker product of Arikan's kernel $T_2=[\begin{smallmatrix} 1&1\\ 1&0 \end{smallmatrix}]$. It can be seen from the definition that $G$ is constructed in a recursive way. As a result, a polar code of size $N$ can be constructed by concatenating two codes of size $N/2$. 

\subsection{MK Polar Codes}
By exploiting $T_2$ as the only kernel of the generator matrix, the block lengths of larger codes will be limited to powers of $2$. However, by considering LDPC WiMAX \cite{Shin2012} code lengths as our guideline we can find out that the code lengths constructed by kernels other than $T_2$ are needed. Utilizing only one or few non-binary kernels however, provides most of the desired code lengths. 
A series of Kronecker products between various kernels can construct the generator matrix as 
\begin{equation}
    G \triangleq T_{n_0}\otimes T_{n_1}\otimes\ldots\otimes  T_{n_s}
    \label{eq:genmat}
\end{equation}
for a code of size $N = n_0\times n_1\times\ldots\times n_s$ with $n_i$s ($0 \leq i \leq s$) being not necessarily individual prime numbers and $T_{n_i}$s being squared matrices.
Each distinctive prime number can be exploited as a kernel's dimensions. However, the least complex and most desirable non-binary kernel, ternary kernel, is defined as $T_3=[\begin{smallmatrix} 1&1&1 \\ 1&0&1\\ 0&1&1  \end{smallmatrix}]$ \cite{gabry2017}. The polarization optimality of $T_3$ is proved in \cite{benammar2017}, though it has a lower polarization exponent compared to $T_2$. In this paper, we investigate codes constructed by any combination of binary and ternary kernels which translates to pure-binary (Arikan's), pure-ternary, and binary-ternary mixed polar codes. 

\begin{figure*}
    \centering
    \includegraphics[width=1.5\columnwidth]{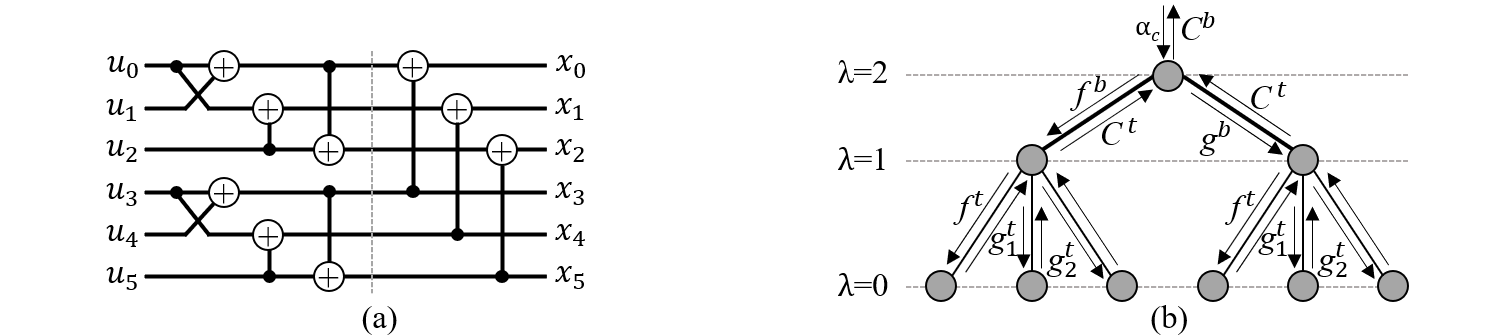}
    \caption{a) Tanner graph and b) decoder tree of a MK polar code of size $N=6$ with $G = T_2 \otimes T_3$.}
    \label{fig:EncDec}
\end{figure*}

The block length of MK codes in this paper can be formulated as $N=2^n\cdot 3^m$ with a generator matrix of $G =  \otimes_{i=0}^{n+m} T_{n_i}$ where $n, m \in \mathbb{N}$ and $0 \leq i \leq n+m$. As an example, we consider the simplest MK polar code of size $N=6$. There are two possible kernel sequences of $T_2\otimes T_3$ and $T_3\otimes T_2$ which results in two different generator matrices since the Kronecker product is not commutative. Thus, different kernel orderings shape exclusive polar codes with distinctive performance characteristics. The Tanner graph of the MK code of size $N=6$ with $G=T_2\otimes T_3$ is illustrated in Fig. \ref{fig:EncDec} (a).
\subsection{Arikan's and MK Successive-Cancellation Decoding}
The SC decoding algorithm is originally proposed by Arikan in \cite{Arikan}. This algorithm can be expanded to decode MK polar codes. The decoder tree of Fig. \ref{fig:EncDec} (b) corresponds  to the Tanner graph of Fig. \ref{fig:EncDec} (a). The soft information ($\alpha_c$) enters the root of the tree in form of LLRs. With the condition of visiting the left node first, the LLRs need to traverse the tree and visit all leaves sequentially so that a codeword can be estimated. To this end, three functions are required in a given node $\nu$. The first one is 
$\alpha_{v_l}$ which is the set of LLRs to be transferred to the left branch and can be computed as 
\begin{equation}
    \alpha^b_{v_l}[i]=sgn(\alpha_v[i] \cdot \alpha_v[i + 2^{(\lambda\text{-}1)}]) min(|\alpha_v[i]| , |\alpha_v[i + 2^{(\lambda\text{-}1)}]|) ~ 
    \label{eq:sc_l}
\end{equation}
where $i \in [0 : 2^{(\lambda\text{-}1)}-1]$ and $\lambda \in [0,n]$ is the level of node $\nu$ in the binary tree. After estimating the hard decisions of the left branch at node $\nu$ ($\beta^b _{v_l}$), we can calculate the set of LLRs to be sent to the right branch using 
\begin{equation}
    \alpha^b_{v_r}[i]= (1-2\beta^b_{v_l}[i])\alpha_v[2i] + \alpha_v[2i+1]~ \textrm{for} ~i \in [0:2^{(\lambda\text{-}1)}-1]
    \label{eq:sc_r}
\end{equation}
where $\alpha^b_{v_r}$ is the LLR of the right branch. After estimating the hard decision bits at the left and right branches, we can combine them to calculate the hard decisions corresponding to node $\nu$ by
\begin{equation}
\begin{aligned}
\relax[\beta^{\nu b}_i, \beta^{\nu b}_{i + 2^{(\lambda\text{-}1)}}] ={} & [\beta^{\nu bl}_i\oplus\beta^{\nu br}_i,\beta^{\nu br}_i],
\label{eq:betak2}
\end{aligned}
\end{equation}
where $\oplus$ represents addition over $\mathbb{F}_2$. In case $\nu$ is a leaf node, the hard decision can be estimated as 
\begin{equation}
\begin{aligned}
\beta_v =     \begin{cases}
h(\alpha _v), & \text{if}\ v \in \mathcal{I} , \\
0, & \text{if}\ v \in \mathcal{F} 
\end{cases},\
h(x) =     \begin{cases}
0,   & \text{if}\ x \geq 0, \\
1,   & \text{otherwise.}
\end{cases}
\label{HD}
\end{aligned}
\end{equation}
Throughout the paper, we refer to (\ref{eq:sc_l}), (\ref{eq:sc_r}) and (\ref{eq:betak2}) as $f^b$, $g^b$ and $C^b$, respectively, as shown in Fig. \ref{fig:EncDec} (b). There are also two general functions that need to be defined. A binary sign function $s(x)$ and a frozen bit indicator vector $a$ formulated as 
\begin{equation}
\begin{aligned}
s(x) =     \begin{cases}
0,   & \text{if}\ l \geq 0, \\
1,   & \text{otherwise}
\end{cases},\
a_i =     \begin{cases}
0,   & \text{if}\ i \in \mathcal{F}, \\
1,   & \text{if}\ i \in \mathcal{I}.
\end{cases}
\label{sx}
\end{aligned}
\end{equation}
A precise pseudo-code of binary SC decoding algorithm is given in the Algorithm \ref{Alg:orgSCk2}.
\begin{algorithm}
$N = lengh(\alpha)$\\
\eIf{$N == 2$}
{   $\beta_0 \gets s(f^b(\alpha))\cdot a(0)$\\
    $\beta_1 \gets s(g^b(\alpha, \beta_0))\cdot a(1)$\\
    return $\beta \gets (\beta_0, \beta_1)$
}{
$\alpha^\prime \gets f^b_{N/2}(\alpha)$\\
$a^\prime \gets a(0~to~N/2\text{-}1)$\\
$\beta^\prime \gets Decode(\alpha^\prime, a^\prime)$\\
$v^\prime \gets C^b(\beta^\prime(0~to~N/4\text{-}1), \beta^\prime(N/4~to~N/2\text{-}1))$\\

$\alpha^{\dblprime} \gets g^b_{N/2}(\alpha, v^\prime)$\\
$a^{\dblprime} \gets a(N/2~to~N\text{-}1)$\\
$\beta^{\dblprime} \gets Decode(\alpha^{\dblprime}, a^{\dblprime})$\\
$v^{\dblprime} \gets C^b(\beta^{\dblprime}(0~to~N/4\text{-}1), \beta^{\dblprime}(N/4~to~N/2\text{-}1))$\\
return $\beta^b \gets C^b(v^\prime, v^{\dblprime})$}
\caption{$\beta^b$ = Decode(\emph{$\alpha$},~\emph{$a$}) using binary SC}
\label{Alg:orgSCk2}
\end{algorithm}

In the case of a pure-ternary node $\nu$ located at level $\lambda$, four functions need to be defined to meet the message-passing criterion. The decoding functions corresponding to the left, middle and right branches are shown by $\alpha^t_{v_l}$, $\alpha^t_{v_c}$ and $\alpha^t_{v_r}$, respectively. For $i \in ~[0, 3^{\lambda{\text -}1}{\text -}1]$ the $\alpha^t_{v_l}$ is calculated as 
\begin{equation}
\begin{aligned}
\alpha^t_{v_l}[i] ={} & \text{sgn}(\alpha_v[i] \cdot \alpha_v[i + 3^{(\lambda\text{-}1)}]\cdot \alpha_v[i + 2\times3^{(\lambda\text{-}1)}]) \\
    & \text{min}(|\alpha_v[i]|, |\alpha_v[i + 3^{(\lambda\text{-}1)}]|, |\alpha_v[i + 2\times3^{(\lambda\text{-}1)}]|).
\label{eq:LLRl}
\end{aligned}
\end{equation}
After calculating the hard decisions from the left branch ($\beta^t_{v_l}$), the LLRs can travel to the middle branch using
\begin{equation}
\begin{aligned}
\alpha^t_{v_c}[i] ={} & (1{\text -}2\beta^t_{v_l}[i])\alpha[i] + f^b(\alpha[i + 3^{(\lambda\text{-}1)}], \alpha[i + 2\times3^{(\lambda\text{-}1)}]).
\label{eq:LLRc}
\end{aligned}
\end{equation}
Finally, using the hard decisions from the left and middle branches, the LLR vector can proceed to the right branch by
\begin{equation}
\begin{aligned}
\alpha^t_{v_r}[i] ={} & (1{\text -}2\beta^t_{v_l}[i])\alpha[i + 3^{(\lambda\text{-}1)}] + (1{\text -}2\beta^t_{v_l}[i]\oplus\beta^t_{v_c}[i])\alpha[i + 2\\\times3^{(\lambda\text{-}1)}].
\label{eq:LLRr}
\end{aligned}
\end{equation}
Now, the hard decisions can be combined at node $\nu$ using
\begin{equation}
\begin{aligned}
\relax[\beta^{\nu t}_i, \beta^{\nu t}_{i + 3^{(\lambda\text{-}1)}},\beta^{\nu t}_{i + 2\times3^{(\lambda\text{-}1)}}] ={} & [\beta^{\nu t_l}_{i}\oplus\beta^{\nu t_c}_{i}, \beta^{\nu t_l}_{i}\oplus\beta^{\nu t_r}_{i}, \beta^{\nu t_l}_{i}\\\oplus\beta^{\nu t_c}_{i}\oplus\beta^{\nu t_r}_{i}].
\label{eq:betak3}
\end{aligned}
\end{equation}
As can be seen in Fig. \ref{fig:EncDec} (b), equations (\ref{eq:LLRl}), (\ref{eq:LLRc}), (\ref{eq:LLRr}) and (\ref{eq:betak3}) are referred as $f^t$, $g_1^t$, $g_2^t$ and $C^t$, respectively. An accurate statement of ternary SC decoding is given in Algorithm \ref{Alg:orgSCk3}. 
\begin{algorithm}
$N = lengh(\alpha)$\\
\eIf{$N == 3$}
{   $\beta_0 \gets s(f^t(\alpha))\cdot a(0)$\\
    $\beta_1 \gets s(g_1^t(\alpha, \beta_0))\cdot a(1)$\\
    $\beta_2 \gets s(g_2^t(\alpha, \beta_0, \beta_1))\cdot a(2)$\\
    return $\beta \gets (\beta_0, \beta_1, \beta_2)$
}{
$\alpha^\prime \gets f^t_{N/3}(\alpha)$\\
$a^\prime \gets a(0~to~N/3\text{-}1)$\\
$\beta^\prime \gets Decode(\alpha^\prime, a^\prime)$\\
$v^\prime \gets C^t(\beta^\prime(0~to~N/3\text{-}1),\beta^\prime(N/3~to~2N/3\text{-}1), \beta^\prime(2N/3~\\to~N\text{-}1))$\\

$\alpha^{\dblprime} \gets g^t_{1N/3}(\alpha, v^\prime)$\\
$a^{\dblprime} \gets a(N/3~to~2N/3\text{-}1)$\\
$\beta^{\dblprime} \gets Decode(\alpha^{\dblprime}, a^{\dblprime})$\\
$v^{\dblprime} \gets C^t(\beta^{\dblprime}(0~to~N/3\text{-}1), \beta^{\dblprime}(N/3~to~2N/3\text{-}1), \beta^{\dblprime}(2N/3\\~to~N\text{-}1))$\\

$\alpha^{\trpprime} \gets g^{t}_{2N/3}(\alpha, v^{\prime}, v^{\dblprime})$\\
$a^{\trpprime} \gets a(2N/3~to~N\text{-}1)$\\
$\beta^{\trpprime} \gets Decode(\alpha^{\trpprime}, a^{\trpprime})$\\
$v^{\trpprime} \gets C^t(\beta^{\trpprime}(0~to~N/3\text{-}1), \beta^{\trpprime}(N/3~to~2N/3\text{-}1), \beta^{\trpprime}(2N/\\3~to~N\text{-}1))$\\
return $\beta^t \gets C^t(v^\prime, v^{\dblprime}, v^{\trpprime})$}
\caption{$\beta^{t}$ = $Decode$($\alpha,~a$) using ternary SC}
\label{Alg:orgSCk3}
\end{algorithm}
\section{MK codes: Construction and Architecture}
\label{sec_arch}
\subsection{MK Code Construction}
\begin{figure*}
    \centering
    \includegraphics[width=2\columnwidth]{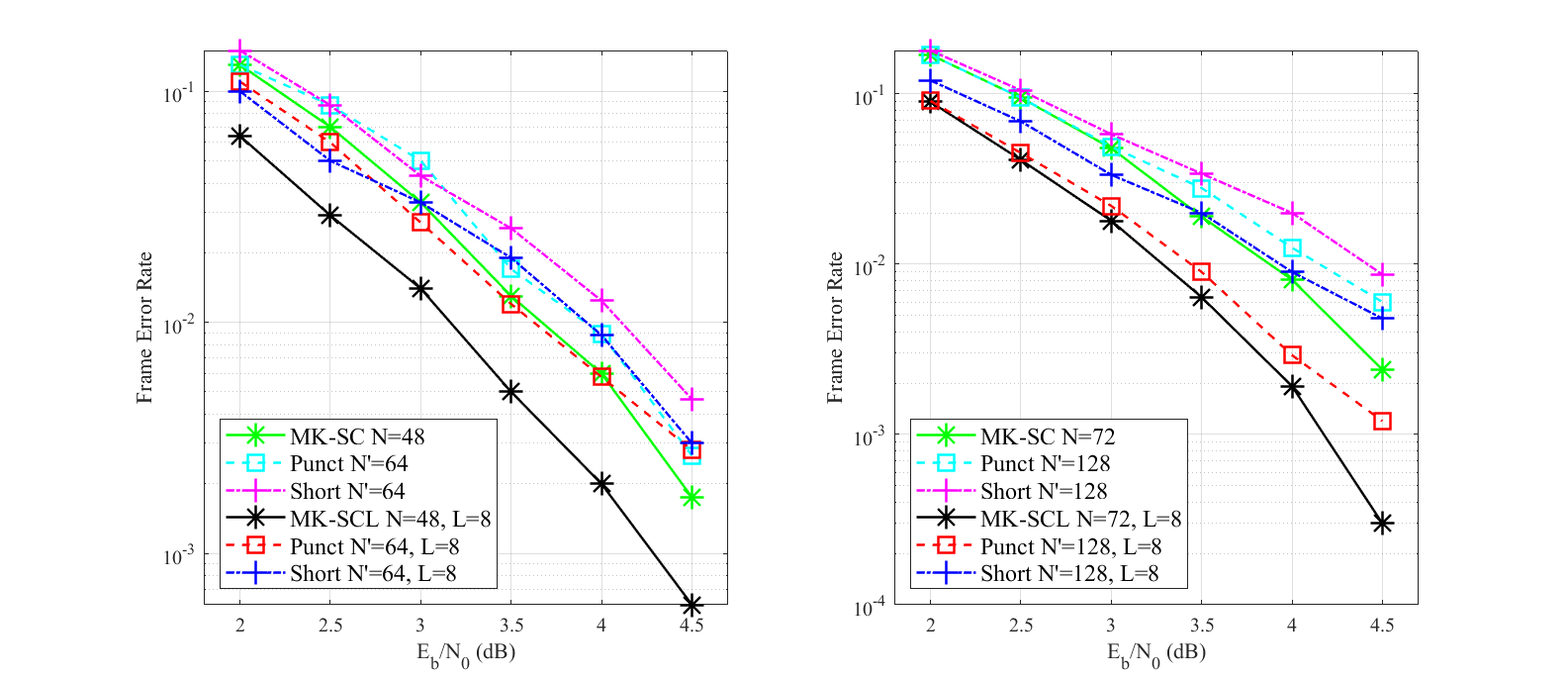}
    \caption{The error-correction performance of MK code of $\mathcal{PC}(72,36)$ compared to that of puncturing and shortening methods.}
    \label{fig:MKCompSCL}
\end{figure*}

 In \cite{gabry2017} a code construction method for MK polar codes which yields significant error-correction performance improvement compared to puncturing \cite{Niu} and shortening \cite{Wang} methods is proposed. The encoding complexity of this method remains low and the same general structure of Arikan's polar codes can be used for decoding. The error-correction performance of MK polar code of size $N=72$ and $\mathcal{R} = {1/2}$ with generator matrix as $G = T_3 \otimes T_2 \otimes T_2 \otimes T_2 \otimes T_3$ is depicted in Fig. \ref{fig:MKCompSCL}. A  binary phase-shift keying (BPSK) modulation over an additive white Gaussian noise (AWGN) channel is used in our simulations. Clearly, the performance of the MK code exceeds that of punctured and shortened codes constructed using a mother code of $N^\prime = 128$. 
 
\begin{figure}
    \centering
    \includegraphics[width=1\columnwidth]{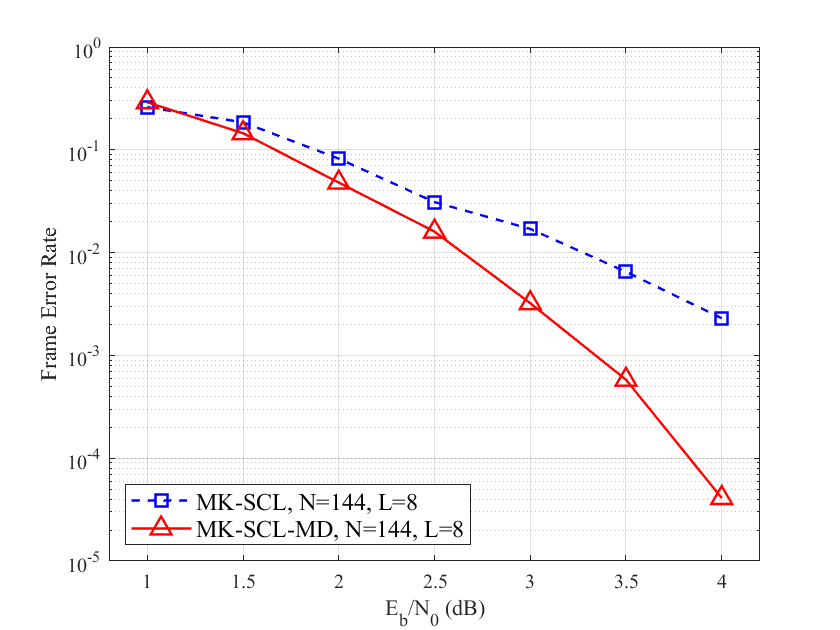}
    \caption{The error-correction performance of MK code of size $\mathcal{PC}(144,72)$ constructed by methods of \cite{gabry2017} (MK-SCL) and \cite{bioglio} (MK-SCL-MD).}
    \label{fig:MKCompSCL2}
\end{figure}

Arbitrary kernel orderings can be employed to construct the MK polar codes. However, given that the Kronecker product is not commutative, different kernel orderings represent distinctive error-correction performance behaviors \cite{gabry2017, bioglio}. At the present time, no theoretical way is available to find the best kernel ordering. Therefore, for a given code we need to perform simulations to find the kernel order offering the best error-correction performance. Throughout this paper, the method proposed in \cite{bioglio} is used to find the kernel orderings since it outperforms \cite{gabry2017} in terms of error-correction performance as depicted in Fig. \ref{fig:MKCompSCL2}. 
As mentioned earlier, the LDPC WiMAX block lengths \cite{Shin2012} are used as our guideline. Thus we mainly focus on the codes desired by this standard. 
\begin{figure*}
    \centering
    \includegraphics[width=2\columnwidth]{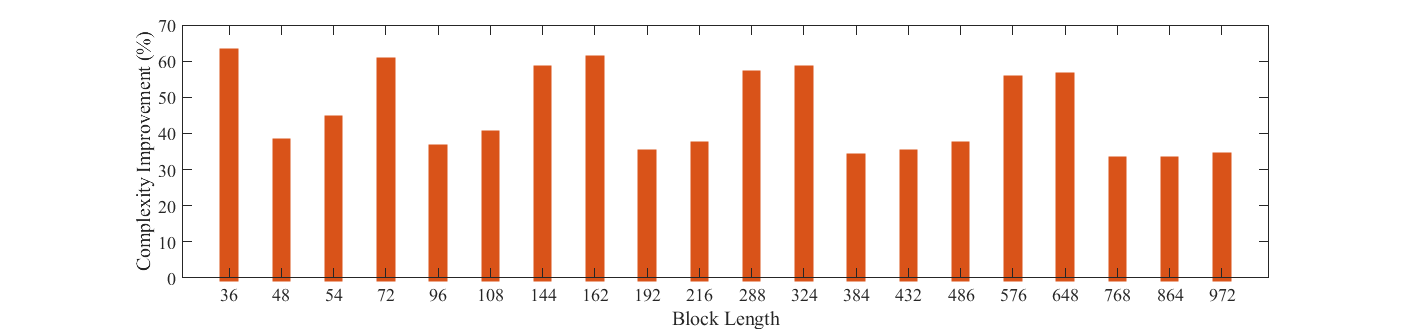}
    \caption{The complexity gain of MK codes versus punctured and shortened codes using the defined complexity metric.}
    \label{fig:CompComparison}
\end{figure*}

In terms of complexity, MK codes feature substantially lower complexity with respect to puncturing and shortening methods. This stems from the fact that the Tanner graph of MK polar codes is smaller in comparison to that of puncturing and shortening methods which employ a mother code of size $N^\prime =2^{\lceil \log_2N \rceil}$. The mother code determines the complexity of punctured and shortened codes. A complexity metric can be defined as the overall required number of LLR computations in each method. We assume that the number of kernels used in the code construction is $s$ which is equivalent to the number of stages in the code's Tanner graph. Using $N\times s$ and $N^\prime \log_2N^\prime$, we can compute the complexity metric of MK and punctured/shortened codes, respectively. 
Fig. \ref{fig:CompComparison} demonstrates the complexity gain of MK codes with reference to punctured and shortened codes for a variety of block lengths. 
Obviously, MK codes offer lower LLR computational complexity ranging from $32.5\%$ to $62.5\%$ compared to that of punctured and shortened codes.
\subsection{Proposed Decoder Architecture}
Given that the SC algorithm contains no loops, we can design a pure combinational architecture that includes no memory elements between the input and output stages. The prime objective is to obtain high throughput. 
In this section, we primarily describe the method of implementing belief propagation functions. After, the proposed combinational architecture of Arikan's, pure-ternary and binary-ternary mixed polar codes will be described. 
\subsubsection{Belief Propagation functions} 
\begin{figure}
    \centering
    \includegraphics[width=1\columnwidth]{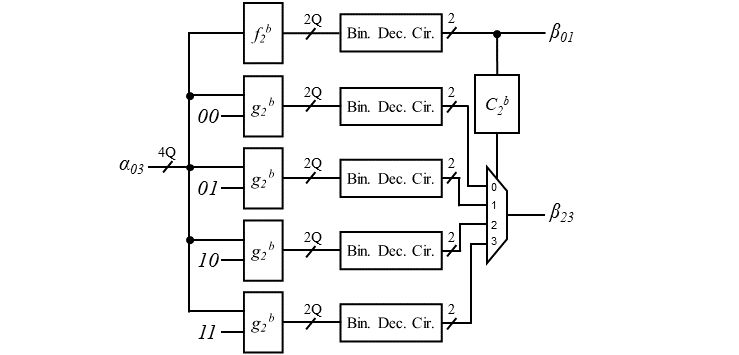}
    \caption{The binary pre-computation circuitry employed in the binary basic building block of size $N=4$.}
    \label{fig:BinPrec}
\end{figure}

\begin{figure}
    \centering
    \includegraphics[width=1\columnwidth]{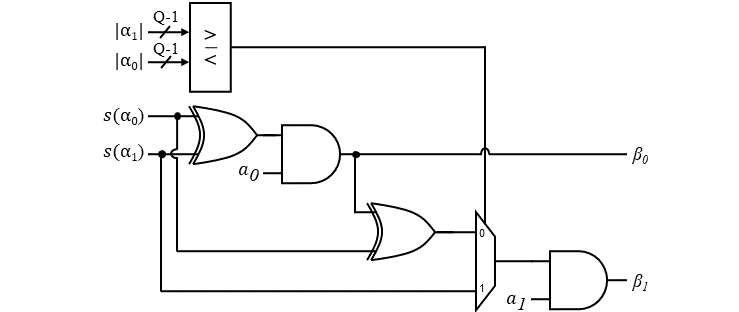}
    \caption{The binary decision logic equivalent to a polar code of size $N=2$.}
    \label{fig:BinDecLog}
\end{figure}

To design the combinational functions, similar to \cite{Dizdar, Leroux} and in order to avoid conversions between different representations, we use $Q$ bits to represent the channel observation LLRs in sign-magnitude format. By directly employing equation (\ref{eq:sc_l}), the $f^b$ function can be implemented by utilizing a comparator and a multiplexer. The pre-computation look-ahead approach can be used at any depth to decrease the latency of SC algorithm \cite{Zhang} at the cost of hardware complexity. Using this technique, all possible output candidates can be pre-computed in one clock cycle, and then the correct candidate can be selected afterward. In all of Arikan's codes of this paper, the polar code of size $N=4$ is used as the basic building block. The $g_2^b$ function in the kernel is implemented by exploiting the pre-computation method. The proposed logic for the pre-computation circuitry exploited in the implementation of the binary basic building block of size $N=4$ is depicted in Fig. \ref{fig:BinPrec}. The frozen bit indicators ($a$) are not shown here.
The implementation of the proposed binary decision logic circuit corresponding to a polar code of size $N=2$ using only comparators, multiplexers, and logic gates is illustrated in Fig. \ref{fig:BinDecLog}. 
To further alleviate the latency and complexity of Arikan's polar codes, the decision logic of Fig. \ref{fig:BinDecLog} estimates the indices in forms of $2i$ and $2i\text{+}1$ ($0\leq i<N/2$) as 
\begin{equation}
\begin{aligned}
\beta_{2i}= (s(\alpha_{2i}) \oplus s(\alpha_{2i+1}))\cdot a_{2i},
\label{u0even}
\end{aligned}
\end{equation}
\begin{equation}
\begin{aligned}
\beta_{2i+1}=     \begin{cases}
s(\alpha_{2i+1})\cdot a_{2i+1}   & \text{if } |\alpha_{2i+1}| \geq |\alpha_{2i}|\\
(s(\alpha_{2i}) \oplus  \beta_{2i})\cdot a_{2i+1}  & \text{otherwise,}
\end{cases}
\label{uodd}
\end{aligned}
\end{equation} 

The pseudo-code of the proposed SC decoding of binary polar codes is summarized in Algorithm \ref{Alg:propSCk2}. There are two key differences between the proposed algorithm in comparison to that of \cite{Dizdar}. The first one is the way we implement the decision logic. Second, the architecture of \cite{Dizdar} exploits an encoder of size $N/2$ in each stage as a part of the glue logic, whereas our proposed architecture employs $C^b$ instead. This leads to consuming substantially lower XOR gates. This fact translates directly into lower latency as $C^b$ is composed of only one layer of XOR gates which has considerably lower latency compared to an encoder of size $N/2$. For instance, in a polar code of size $N=32$, $42,3\%$ of the overall XOR gates can be saved. This modification also results in reducing the consumption of the interconnect resources. Saving interconnect resources is important because interconnect congestion is a phenomenon that limits the performance of large combinational circuits implemented on FPGA.

\begin{figure}
    \centering
    \includegraphics[width=1\columnwidth]{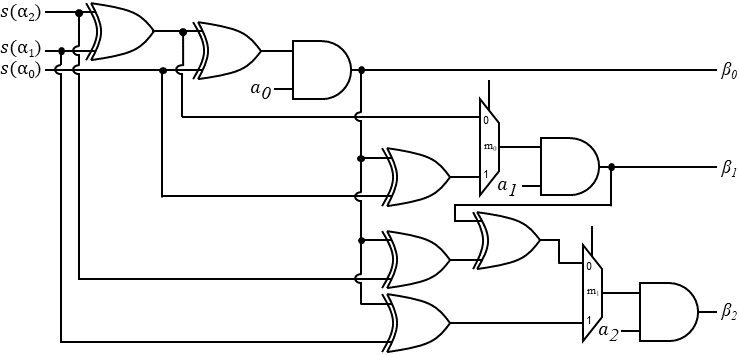}
    \caption{The proposed ternary decision logic equivalent to a polar code of size $N=3$.}
    \label{fig:TerDecLog}
\end{figure}

\begin{figure}
    \centering
    \includegraphics[width=1\columnwidth]{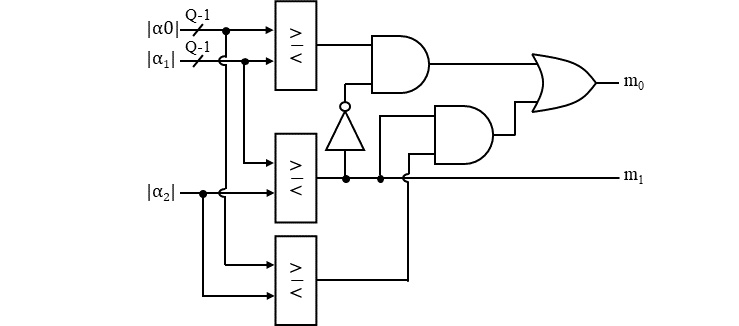}
    \caption{The proposed control logic of the ternary decision logic of Fig. \ref{fig:TerDecLog}.}
    \label{fig:TerBuildingBlockCTRL}
\end{figure}

\begin{algorithm}
$N = lengh(\alpha)$\\
\eIf{$N == 2$}{
    $\beta_0 \gets (s(\alpha(0)) \oplus s(\alpha(1)))\cdot a(0)$\\
    \eIf{$abs(\alpha(1)) \geq abs(\alpha(0))$}{
    $\beta_1 \gets s(\alpha(1))\cdot a(1)$}{
    $\beta_1 \gets (s(\alpha(0)) \oplus \beta(0))\cdot a(1)$}
    return $\beta \gets (\beta_0, \beta_1)$
}{
\eIf{$N == 4$}
{   $\alpha^\prime \gets f^b_{2}(\alpha)$\\
    $a^\prime \gets a(0~to~1)$\\
    $\beta^\prime \gets Decode(\alpha^\prime, a^\prime)$\\
    $v^\prime \gets C^b(\beta^\prime(0), \beta^\prime(1))$\\
    
    $\alpha_0^{\dblprime} \gets g^b_{2}(\alpha, (0,0))$\\
    $\alpha_1^{\dblprime} \gets g^b_{2}(\alpha, (0,1))$\\
    $\alpha_2^{\dblprime} \gets g^b_{2}(\alpha, (1,0))$\\
    $\alpha_3^{\dblprime} \gets g^b_{2}(\alpha, (1,1))$\\
    $a^{\dblprime} \gets a(2~to~3)$\\
    
    $\beta_0^{\dblprime} \gets Decode(\alpha_0^{\dblprime}, a^{\dblprime})$\\
    $\beta_1^{\dblprime} \gets Decode(\alpha_1^{\dblprime}, a^{\dblprime})$\\
    $\beta_2^{\dblprime} \gets Decode(\alpha_2^{\dblprime}, a^{\dblprime})$\\
    $\beta_3^{\dblprime} \gets Decode(\alpha_3^{\dblprime}, a^{\dblprime})$\\
    
    \eIf{$v^\prime(0) == 0$}{\eIf{$v^\prime(1) == 0$}{
    $v^{\dblprime} \gets C^b(\beta^{\dblprime}_{0}(0), \beta^{\dblprime}_{0}(1))$
    }{$v^{\dblprime} \gets C^b(\beta^{\dblprime}_{1}(0), \beta^{\dblprime}_{1}(1))$)}}{
    \eIf{$v^\prime_{1} == 0$}{$v^{\dblprime} \gets C^b(\beta^{\dblprime}_{2}(0), \beta^{\dblprime}_{2}(1))$)}{
    $v^{\dblprime} \gets C^b(\beta^{\dblprime}_{3}(0), \beta^{\dblprime}_{3}(1))$)}}
    return $\beta \gets (v^\prime, v^{\dblprime})$
}{
$\alpha^\prime \gets f^b_{N/2}(\alpha)$\\
$a^\prime \gets a(0~to~N/2\text{-}1)$\\
$\beta^\prime \gets Decode(\alpha^\prime, a^\prime)$\\
$v^\prime \gets C^b(\beta^\prime(0~to~N/2\text{-}1), \beta^\prime(N/2~to~N\text{-}1))$\\

$\alpha^{\dblprime} \gets g^b_{N/2}(\alpha, v^\prime)$\\
$a^{\dblprime} \gets a(N/2~to~N\text{-}1)$\\
$\beta^{\dblprime} \gets Decode(\alpha^{\dblprime}, a^{\dblprime})$\\
$v^{\dblprime} \gets C^b(\beta^{\dblprime}(0~to~N/2\text{-}1), \beta^{\dblprime}(N/2~to~N\text{-}1))$\\
return $\beta \gets C^b(v^\prime, v^{\dblprime})$}}
\caption{$\beta^b$ = $Decode$(\emph{$\alpha$},~\emph{$a$}) using the proposed approach}
\label{Alg:propSCk2}
\end{algorithm}
In order to implement the combinational ternary decoder, three functions $f^t$, $g_1^t$ and $g_2^t$ are needed to be implemented. Similar to $f^b$ and using equation (\ref{eq:LLRl}), we can implement the $f^t$ function by exploiting two comparators and a multiplexer. To further reduce the latency and complexity of the ternary decoder, we estimate the codewords with indices in forms of $3i$, $3i\text{+}1$ and $3i\text{+}2$ ($0\leq i<N/3$) as
\begin{equation}
\begin{aligned}
\beta_{3i}= (s(\alpha_{3i}) \oplus s(\alpha_{3i+1}) \oplus s(\alpha_{3i+2}))\cdot a_{3i},
\label{HD0}
\end{aligned}
\end{equation}
\begin{equation}
\begin{aligned}
\beta_{3i+1}=     \begin{cases}
(s(\alpha_{3i})\oplus \beta_{3i})\cdot a_{3i+1}   & \text{if } (|\alpha_{3i}| \geq |\alpha_{3i+1}|\\& \text{and }\\& |\alpha_{3i+2}| \geq |\alpha_{3i+1}|) 
    \\&\text{or } (|\alpha_{3i}| \geq |\alpha_{3i+2}|\\& \text{and }\\& |\alpha_{3i+1}| \geq |\alpha_{3i+2}|) \\

(s(\alpha_{3i+1}) \oplus s(\alpha_{3i+2}))\cdot a_{3i+1}   & \text{otherwise,}
\end{cases}
\label{HD1}
\end{aligned}
\end{equation}
\begin{equation}
\begin{aligned}
\beta_{3i+2}=     \begin{cases}
(s(\alpha_{3i+1}) \oplus \beta_{3i})\cdot a_{3i+2} & \text{if } |\alpha_{3i+1}| \geq\\& |\alpha_{3i+2}|  \\
(s(\alpha_{3i+2}) \oplus \beta_{3i} \oplus \beta_{3i+1})\cdot a_{3i+2}  & \text{otherwise,}
\end{cases}
\label{HD2}
\end{aligned}
\end{equation}
where $\alpha_0$, $\alpha_1$, and $\alpha_2$ are the input LLRs to $g_1^t$, and $g_2^t$ functions and $a_0$, $a_1$ and $a_2$ are the frozen bit indicators corresponding to each input LLR. Fig. \ref{fig:TerDecLog} portrays the proposed ternary decision logic implemented by multiplexers and logic gates. The control logic is separately depicted in Fig. \ref{fig:TerBuildingBlockCTRL} to prevent congestion. The control logic generates the required signals for controlling two multiplexers of Fig. \ref{fig:TerDecLog} ($m_0$ and $m_1$) implemented by only using comparators and logic gates.
Algorithm \ref{Alg:propSCk3} gives the pseudo-code of the proposed ternary SC decoding algorithm.
\begin{algorithm}
$N = lengh(\alpha)$\\
\eIf{$N == 3$}
{   $\beta_0 \gets (s(\alpha(0)) \oplus s(\alpha(1)) \oplus s(\alpha(2))).a(0)$\\
    \eIf{$abs(\alpha(0)) \geq abs(\alpha(1))$ and $abs(\alpha(2)) \geq abs(\alpha(1))$ 
    or $abs(\alpha(0)) \geq abs(\alpha(2))$ and $abs(\alpha(1)) \geq abs(\alpha(2))$ }{
    $\beta_1 \gets (s(\alpha(0)) \oplus \beta_0)\cdot a(1)$ }{
    $\beta_1 \gets (s(\alpha(1)) \oplus s(\alpha(2)))\cdot a(1)$ }
    \eIf{$abs(\alpha(1)) \geq abs(\alpha(2))$}{
    $\beta_2 \gets (s(\alpha(1)) \oplus \beta_0)\cdot a(2)$}{
    $\beta_2 \gets (s(\alpha(2)) \oplus \beta_0 \oplus \beta_1)\cdot a(2)$}
    return $\beta \gets (\beta_0, \beta_1, \beta_2)$
}{
$\alpha^\prime \gets f^t_{N/3}(\alpha)$\\
$a^\prime \gets a(0~to~N/3\text{-}1)$\\
$\beta^\prime \gets Decode(\alpha^\prime, a^\prime)$\\
$v^\prime \gets C^t(\beta^\prime(0~to~N/3\text{-}1),\beta^\prime(N/3~to~2N/3\text{-}1), \beta^\prime(2N/3~to~\\N\text{-}1))$\\

$\alpha^{\dblprime} \gets g^t_{1N/3}(\alpha, v^\prime)$\\
$a^{\dblprime} \gets a(N/3~to~2N/3\text{-}1)$\\
$\beta^{\dblprime} \gets Decode(\alpha^{\dblprime}, a^{\dblprime})$\\
$v^{\dblprime} \gets C^t(\beta^{\dblprime}(0~to~N/3\text{-}1), \beta^{\dblprime}(N/3~to~2N/3\text{-}1), \beta^{\dblprime}(2N/3~\\~to~N\text{-}1))$\\

$\alpha^{\trpprime} \gets g^t_{2N/3}(\alpha, v^{\prime}, v^{\dblprime})$\\
$a^{\trpprime} \gets a(2N/3~to~N\text{-}1)$\\
$\beta^{\trpprime} \gets Decode(\alpha^{\trpprime}, a^{\trpprime})$\\
$v^{\trpprime} \gets C^t(\beta^{\trpprime}(0~to~N/3\text{-}1), \beta^{\trpprime}(N/3~to~2N/3\text{-}1), \beta^{\trpprime}(2N\\/3~to~N\text{-}1))$\\
return $\beta^t \gets C^t(v^\prime, v^{\dblprime}, v^{\trpprime})$}
\caption{$\beta^t$ = Decode(\emph{$\alpha$},~\emph{$a$}) using the proposed approach}
\label{Alg:propSCk3}
\end{algorithm}
\subsubsection{Overall Architecture} 
\begin{figure}
    \centering
    \includegraphics[width=1\columnwidth]{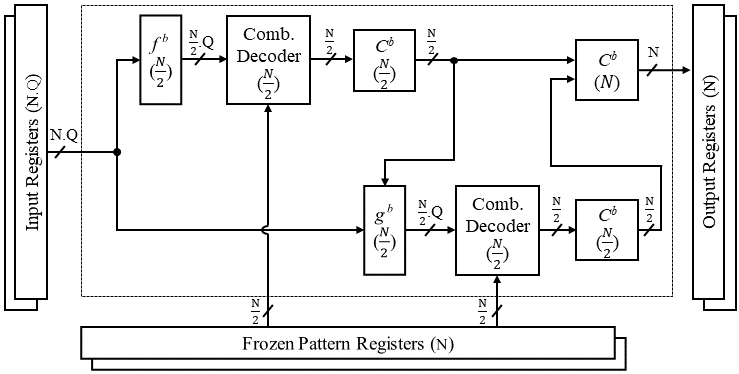}
    \caption{The proposed combinational architecture of the Arikan's decoder.}
    \label{fig:K2Datapath}
\end{figure}

In this section, we propose the hardware architectures corresponding to Algorithms \ref{Alg:propSCk2} and \ref{Alg:propSCk3}.
Fig. \ref{fig:K2Datapath} depicts the generalized combinational architecture of a decoder of size $N$ constructed using Arikan's kernel. An Arikan's decoder of size $N$ is composed of two decoders of size $N/2$ glued by a $f^b$, a $g^b$, and a $C^b$. 
A decoder of size $N=4$ is exploited as the basic building block of Arikan's decoders. Unlike the decoder of \cite{Dizdar}, the proposed decoder includes the required memory to load the next LLR frame and frozen bit indicator set, and also the necessary memory to offload the earlier estimated codeword. 

\begin{figure}
    \centering
    \includegraphics[width=1\columnwidth]{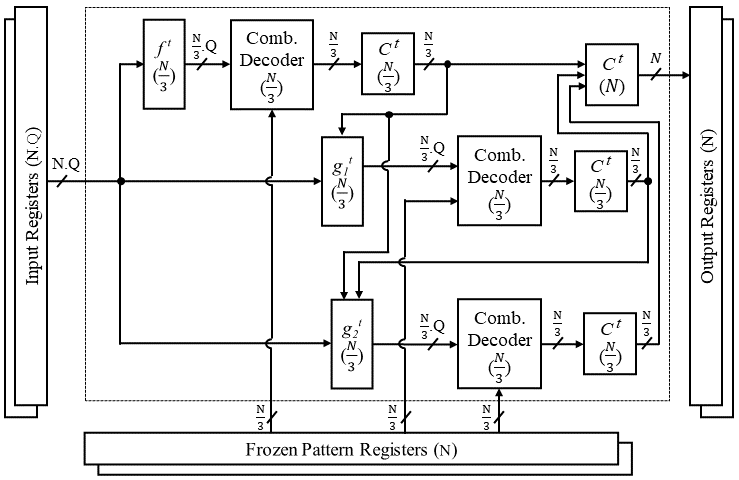}
    \caption{The proposed combinational architecture of the pure-ternary decoder.}
    \label{fig:K3Datapath}
\end{figure}

Fig. \ref{fig:K3Datapath} illustrates the proposed combinational architecture of a pure-ternary polar code of size $N$. A decoder of size $N$ is constructed by three decoders of size $N/3$. The glue logic includes one $f^t$, one $g_1^t$, one $g_2^t$ and two $C^t$ of size $N/3$. A decoder of size $N=3$ is employed as the basic building block in this case. The memory structure is similar to that of the proposed Arikan scheme. 

As a result of the recursive nature of the SC algorithm, a MK code of size $N$ can be constructed by a mixed design of Arikan's and ternary architectures. The kernel order determines whether to use $N=3$ or $N=4$ as the decision logic circuitry. In case the kernel order does not meet the condition to use $N=4$ as the decision circuitry, it can be replaced by $N=2$. The only difference is that the pre-computation method is no longer used. However, in all MK codes of this paper, the condition for using $N=4$ as the decision circuit is met. After selecting the basic building block, the kernel order determines whether to use a binary or ternary glue circuitry as the top stage. We continue employing the glue logic stages until the target block code is constructed. 

\begin{figure}
    \centering
    \includegraphics[width=1\columnwidth]{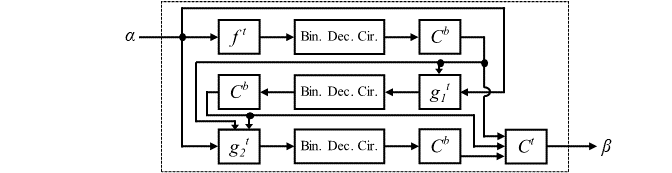}
    \caption{The combinational MK polar decoder for $N=6$ and $G = T_3 \otimes T_2$.}
    \label{fig:N6Datapath}
\end{figure}

The architecture of a MK polar code of size $N = 6$ correlated to the decoder tree of Fig. \ref{fig:EncDec} (b) with $G=T_3\otimes T_2$ is displayed in Fig. \ref{fig:N6Datapath}. To avoid congestion, the registers and frozen bit indicators are not shown here. A binary decision-making circuitry ($N=2$) is exploited as the basic building block of this decoder as the last kernel in the kernel sequence is a $T_2$. Having a $T_3$ as the next kernel, the glue logic is composed of one $f^t$, one $g_1^t$, one $g_2^t$, and three binary combine logics ($C^b$). A ternary combine function ($C^t$) is also employed to receive the estimated codeword at the root of the tree before writing them into the output registers. Whether to use a binary or ternary combine logic before writing the estimated codeword into the output registers is determined by the first kernel in the kernel sequence ($T_3$ here).
\subsubsection{Memory Architecture} 
The proposed Arikan's, pure-ternary and MK combinational architectures occupy $N \times (Q+2)$ register bits. These memory elements are utilized to store the input LLRs ($N \times Q$ bits), estimated codeword (N bits), and frozen bit indicator set (N bits). It can be seen from Fig. \ref{fig:K2Datapath}, Fig. \ref{fig:K3Datapath} and Fig. \ref{fig:N6Datapath} that there are no synchronous logic elements, i.e registers or RAM arrays, between the input and output registers. This characteristic of combinational decoders leads to power and processing time efficiency. Removing the RAM routers also causes decreased hardware complexity and eliminates long read/write latencies. The latency of the decoder is one clock cycle since it generates the estimated codeword in one long clock cycle after accepting the input LLRs. Therefore, the logic between the input and output registers determines the overall critical path. 
\subsection{Complexity Analysis of the MK Decoder}
The complexity of the proposed architecture can be expressed as the total number of the basic building blocks i.e. comparators, adders, and subtractors in the design. 
Let's assume that $c_N^b$ indicates the number of comparators utilized in implementing $f^b$. It can be seen in Algorithm \ref{Alg:propSCk2} that the initial value of the consumed comparators for the Arikan's decoder is $c_4^b=2$.
It is shown in \cite{Dizdar} that the total number of basic building blocks of a combinational-logic-based Arikan's decoder of size $N$ with $c_4^b=2$ can be estimated as 
\begin{equation}
c_N^b+s_N^b+r_N^b = N(\frac{3}{2}\log_2(N)-1)\approx \frac{3}{2}N\log_2(N),    
\label{complex_bin}
\end{equation}
where $s_N^b$ and $r_N^b$ express the number of comparators consumed in the decision logic and the total number of adders and subtractors employed in implementing $g^b$, respectively. Equation (\ref{complex_bin}) proves that the complexity of the Arikan's combinational decoder is in the order of $\mathcal{O}(N\log_2(N))$.

For a pure-ternary polar code, the number of comparators ($c_N^t$) used for implementing $f^t$ and $g_1^t$ for a decoder of size $N$ has the recursive relationship of 
\begin{equation}
c_N^t = 3c_{\frac{N}{3}}^t+N=3(3c_{\frac{N}{9}}^t+{\frac{N}{3}})+N=\ldots.
\label{com_ter}
\end{equation}
Using Algorithm \ref{Alg:propSCk3}, we can initialize $c_N^t$ by $c_3^t=3$. The recursion equation of (\ref{com_ter}) can exactly be computed as $N\log_3(N)$.
Knowing that the number of comparators used in the decision logic for $N=3$ is $s_3^t = 3$, we obtain $s_N^t = N$. 

Finally, the number of adders and subtractors will be estimated. The function $g_1^t$ is implemented by one adder and one subtractor, and function $g_2^t$ is implemented by two adders and two subtractors. As result, the total number of adders and subtractors can be computed as $r_N^t = \frac{3}{2}c_N^t$.
Thus, the number of basic logic blocks of a pure-ternary decoder can be estimated as 
\begin{equation}
c_N^t+s_N^t+r_N^t = N(\frac{5}{2}\log_3(N)+1)\approx \frac{5}{2}N\log_3(N),
\label{complex_ter}
\end{equation}
Equation (\ref{complex_ter}) verifies that the complexity of the pure-ternary combinational decoder is in the order of $\mathcal{O}(N\log_3(N))$. Using (\ref{complex_bin}) and (\ref{complex_ter}), it is obvious that compared Arikan's and pure-ternary polar codes with block lengths in the same range (take $N^b=2048$ and $N^t=2187$ as an example), the complexity of Arikan's codes is lower than that of pure-ternary codes. This is due to the fact that ternary belief propagation functions and decision logic circuits are more complex than those of Arikan. Therefore, we can conclude that the complexity of the proposed mixed-kernel decoders with $N_{min}=2$ and $N_{max}=4096$ is lower-bounded by $\frac{3}{2}N\log_2(N)$ and upper-bounded by $\frac{5}{2}N\log_3(N)$, i.e. 
\begin{equation}
\frac{3}{2}N\log_2(N) \leq c_N^{MK} \leq \frac{5}{2}N\log_3(N).
\label{complex_MK}
\end{equation}
It should be noted that providing a general equation for complexity analysis of MK codes is not possible as it directly depends on the number and location of different kernels in the kernel sequence.
\section{Auto-Generation of Combinational Polar Decoders}
\label{sec_compiler}
\subsection{High-Level Synthesis}
The process of transforming a higher-level description of algorithms or behaviors into a register-transfer level (RTL) is known as high-level synthesis (HLS) \cite{nane2015survey}. HLS tools transform high-level programming languages such as C/C++ or Python into hardware description language (HDL). In MK combinational architecture, all the HDL sub-modules need to be modified when the block length or the kernel order is changed. The proposed algorithms can automatically be transformed into HDL to speed up the development process of combinational decoders.
\subsection{Generation Process}
Using Algorithms \ref{Alg:propSCk2} and \ref{Alg:propSCk3}, we developed a polar compiler \cite{PythonCode} scripted in Python to automate the process of generating HDL files for implementing various size decoders with different kernel orderings. 
The user needs to enter the block length and optional kernel order of the target polar code. In case the user does not enter a kernel order, the compiler automatically assigns the kernel ordering with the highest error-correction performance. There are several functions corresponding to formulas (\ref{eq:sc_l}-\ref{eq:betak3}), basic building blocks, top modules and sub-modules, and interface. The polar compiler calls relevant functions based on the predefined rules. The functions take their inputs from the compiler's top module and output the requested VHDL files. Generally, the process of compilation is similar to HLS flow: 

\setlength{\parindent}{0pt} Parameters (specified by user) $\rightarrow$ functions (high-level description) $\rightarrow$ VHDL files (HDL)

\subsection{Time Efficiency}
\begin{figure*}
    \centering
    \includegraphics[width=2\columnwidth]{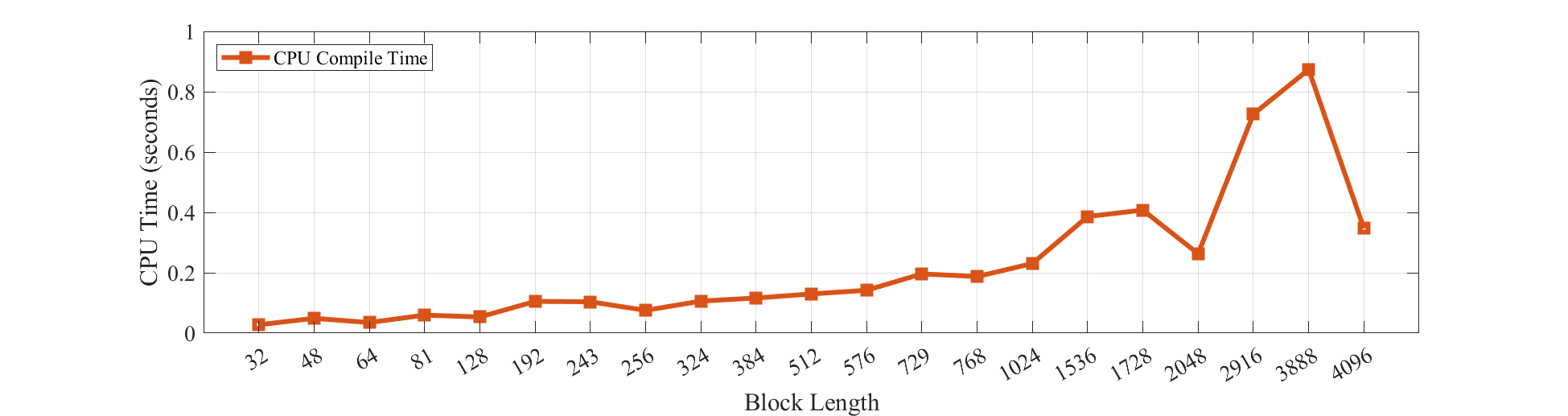}
    \caption{Compile time for various size polar codes.}
    \label{fig:RunTime}
\end{figure*}

The time efficiency of the proposed polar compiler is evaluated using an AMD Ryzen 7 PRO 5850U x64 CPU operating at 1.90 GHz frequency. The required time for generating all necessary VHDL files for polar decoders of various sizes are shown in Fig. \ref{fig:RunTime}. Each data value is measured by running the proposed compiler $20$ times and calculating the average value. The CPU time changes based on the required number and complexity of sub-modules which is directly affected by the block length as well as the ordering of kernels. Therefore, it is expected that the required time for compiling MK decoders is higher than that of Arikan's decoders as can be seen in Fig. \ref{fig:RunTime}. However, for the majority of polar codes, the compile time is less than $0.4$ seconds whereas that of the longest MK polar code is $0.88$ seconds. Thus using the proposed polar compiler is an efficient way of generating all the required VHDL files for combinational polar decoders. 
\section{Implementation Results and Comparison}
\label{sec_res}
All polar codes of this paper are described using VHDL coding in Xilinx Vivado 2019.1 environment. In order to validate the design, logic synthesis, technology mapping, and place-and-route are conducted targeting Xilinx Virtex-6 FPGA (40 nm). Using BPSK modulation over an AWGN channel, a software program generates the random codewords and transfers them to the FPGA. The flexibility and scalability of the proposed decoder are evaluated by implementing different codes with different kernel orderings. In our experiment, we use an extra set of registers to store the input, output, and frozen pattern data. This method allows the decoder to decode a frame with a given frozen pattern, and load another frame and its corresponding frozen set which prevents performance degradation. Likewise, the estimated codeword can be transferred while another decoding is ongoing. 

To facilitate the comparison between different schemes, the decoding latency is defined as the time required for decoding a frame. Similar to the binary case, the coded and information throughput of MK polar codes can be calculated as $\mathcal{T_C}=N.f$ and $\mathcal{T_C}=N.f.\mathcal{R}$, respectively. 

\subsection{Error-Correction Performance and Quantization}
\begin{figure*}
    \centering
    \includegraphics[width=2\columnwidth]{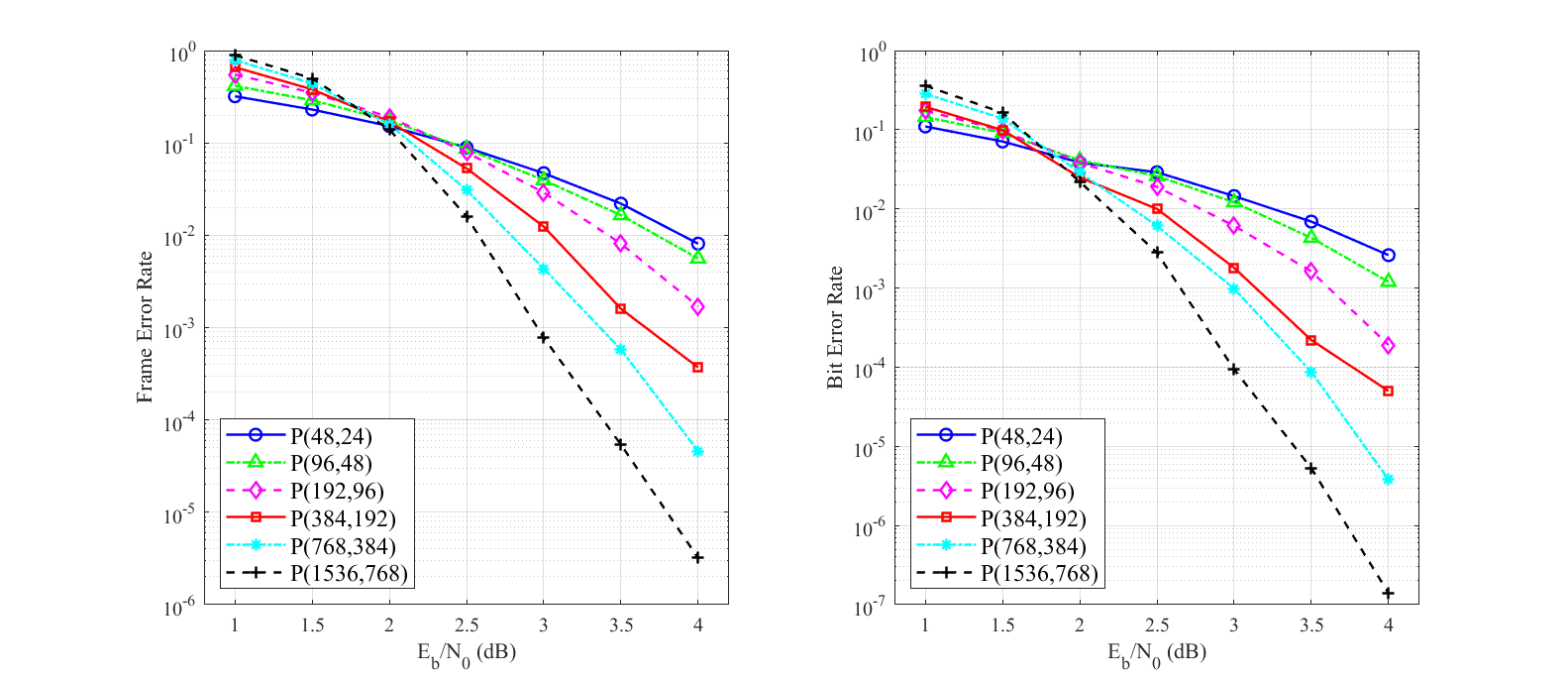}
    \caption{The error-correction performance of MK polar codes of rate $R = \frac{1}{2}$.}
    \label{fig:MKECC}
\end{figure*}

As mentioned earlier, the LDPC WiMAX standard \cite{Shin2012} states that a considerable number of block lengths can be constructed by using only one or a few non-binary kernels. The error-correction performance of codes exploiting only one non-binary kernel and representing LLRs in floating-point format is depicted in Fig. \ref{fig:MKECC} as this standard pays special attention to such codes.

\begin{figure*}
    \centering
    \includegraphics[width=2\columnwidth]{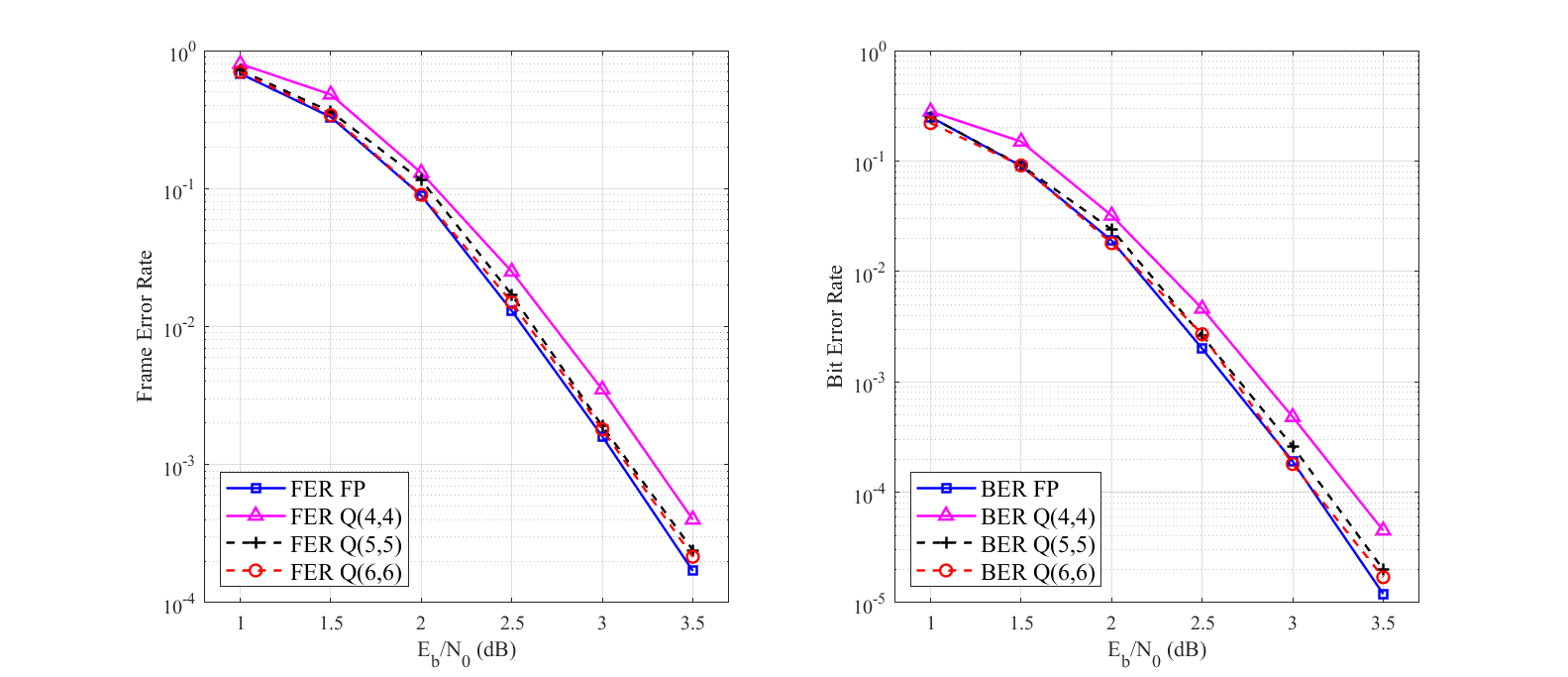}
    \caption{Impact of LLR quantization on the error-correction performance of MK code of $\mathcal{PC}(1024,512)$.}
    \label{fig:MKQunt}
\end{figure*}

We define the quantization scheme as $Q(Q_i, Q_c)$, where $Q_i$ and $Q_c$ stand for the total number of bits used for representing the number of internal and channel bits of LLRs, respectively. Fig. \ref{fig:MKQunt} compares the performance loss of $Q(4, 4)$, $Q(5, 5)$, and $Q(6, 6)$ for a polar code of $\mathcal{PC}(1024,512)$. Obviously, $Q(5, 5)$ leads to an error-correction performance fairly close to that of the floating-point counterpart with a very negligible margin compared to $Q(6, 6)$. Therefore in this work, we select  $Q(5, 5)$ as the quantization scheme.

\subsection{FPGA Implementation Results and Comparison} 
Routing in a combinational decoder gets more complicated as the code length increases due to consuming a larger number of logic blocks. The interconnect delay, therefore, increases as the code length grows. Aspecially this phenomenon reveals itself in FPGAs (as opposed to ASICs) due to using pre-fabricated routing resources. 

Table \ref{tab:utilizationK2} summarizes the FPGA utilization and timing performance of the proposed combinational decoder versus that of \cite{Dizdar} for a wide range of block lengths. It can be seen that in all cases, the proposed decoder roughly doubles the operating frequency and coded throughput. Similar to \cite{Dizdar}, the performance of the proposed decoder drops as the codelength grows due to interconnect delay. However, this effect is lower in our implementation since it consumes lower logic resources (ranges from 16\% to 47\%) which is generally due to replacing the encoders with combine logics in the decoder's architecture. It should be noted that no look-up tables (LUTs) are used as memory. The registers in our design are used for retaining the input LLRs, frozen bit pattern as well as the estimated output. They are also employed for implementing small logic circuits. The proposed combinational decoder consumes from 2.6 to 3.09 times the number of registers in comparison to \cite{Dizdar}. The main reason is that \cite{Dizdar} does not support loading the next frame (and its corresponding frozen pattern bits) while decoding another. It also does not contain the required memory for offloading the previously estimated codeword while another frame is getting decoded. Finally, the proposed decoder does not consume any RAM, while \cite{Dizdar} occupies 206 to 7168 bits of RAM. From Table \ref{tab:utilizationK2}, it can also be seen that the scalability of the decoder is also improved. The resource consummation of a code of length $N$ is 2 times greater than that of a code of $N/2$ plus some overhead stemming from the glue logic that connects two $N/2$ codes.

\begin{table*}
\centering
\caption{Post-fitting results of various Arikan's polar codes}
\begin{tabular}{cccccccccccc}
\hline
 &&&This work&&&&&&\cite{Dizdar}&& \\\cline{2-6}  \cline{8-12} 
\begin{tabular}[c]{@{}c@{}}Block \\Length  \end{tabular} & LUTs  & Registers & \begin{tabular}[c]{@{}c@{}}RAM \\(bits)  \end{tabular}&
\begin{tabular}[c]{@{}c@{}}f \\(MHz)  \end{tabular}&
\begin{tabular}[c]{@{}c@{}}T/P \\(Mbps) \end{tabular}&& LUTs  & Registers & \begin{tabular}[c]{@{}c@{}}RAM \\(bits)  \end{tabular}&
\begin{tabular}[c]{@{}c@{}}f \\(MHz)  \end{tabular}&
\begin{tabular}[c]{@{}c@{}}T/P \\(Mbps) \end{tabular} \\\hline
32&  1611   & 742  & 0&51.875&1660&& 1918& 206&224&27.5&880\\ 
64&  3921   & 1480 & 0&25.625&1640&& 5126&392&448&13,28&850\\ 
128& 8926   & 2985 & 0&11.3&1446.4&& 14517&783&896&6.4&820\\ 
256& 20259  & 6058 & 0&5.34&1367.9&&35152&1561&1792&2.93&750\\ 
512& 45986  & 12334& 0&2.49&1275.7&& 77154&3090&3584&1.43&730\\
1024&102740 & 25139& 0&1.19&1214.2&& 193456&6151&7168&0.59&600\\
\hline
\end{tabular}
\label{tab:utilizationK2}
\end{table*}

The FPGA utilization and performance parameters of various pure-ternary and MK codes are tabulated in Table \ref{tab:utilizationMK}. We have tried to incorporate all possible kernel orderings. The resource consumption is determined by the number of kernels used to construct a given code. Although ternary layers occupy more resources compared to the binary layers, however, they construct bigger codes. For instance, two codes of size 512 (9 kernels) from Table \ref{tab:utilizationK2} and 576 (8 kernels) from Table \ref{tab:utilizationMK} almost occupy the same amount of LUTs and registers. Therefore, the resource consumption is proportional to the target block length, not the kernel order nor the basic building block, which again shows the scalability of the design. Table \ref{tab:utilizationMK} also provides the performance parameters of the same block lengths reported in \cite{Coppolino} for the sake of comparison.
\begin{table*}
\centering
\caption{Post-fitting results of various MK polar codes}
\begin{tabular}{cccccccccccc}
\hline
 &&&&\multicolumn{2}{c}{This work}&&&&&\multicolumn{1}{c}{\cite{Coppolino}}\\\cline{3-8} \cline{10-12} 
\begin{tabular}[c]{@{}c@{}}Block \\Length  \end{tabular}  &Kernel Order& LUTs  & Registers & \begin{tabular}[c]{@{}c@{}}RAM \\(bits)  \end{tabular} &
\begin{tabular}[c]{@{}c@{}}L \\($\mu$s)  \end{tabular}&
\begin{tabular}[c]{@{}c@{}}f \\(MHz)  \end{tabular}&
\begin{tabular}[c]{@{}c@{}}T/P \\(Mbps) \end{tabular}&&
\begin{tabular}[c]{@{}c@{}}L \\($\mu$s)  \end{tabular}&
\begin{tabular}[c]{@{}c@{}}f \\(MHz)  \end{tabular}&
\begin{tabular}[c]{@{}c@{}}T/P \\(Mbps) \end{tabular}\\\hline
48& \{3,2,2,2,2\}         & 3049  & 1088  & 0 &0.029 & 34.17 & 1640   &&0.111 &1230 & 430.9\\ 
81&\{3,3,3,3\}            & 5134  & 2129  & 0 &0.049 & 20.55 & 1664.5 &&0.132 &1230 & 615  \\ 
192&\{3,2,2,2,2,2,2\}     & 15005 & 4426  & 0 &0.12  & 8.3   & 1592.7 &&0.477 &1230 & 402.3\\ 
243&\{3,3,3,3,3\}         & 1997  & 3443  & 0 &0.147 & 6.78  & 1646.5 &&0.422 &1230 & 578.1 \\ 
324&\{2,2,3,3,3,3\}       & 27037 & 8719  & 0 &0.218 & 4.59  & 1487.8 &&0.587 &1110 & 555 \\ 
384&\{3,2,2,2,2,2,2,2\}   & 32575 & 9003  & 0 &0.288 & 3.47  & 1330.9 &&1.04  &1110 & 368.7\\ 
576&\{2,2,2,2,2,2,3,3\}   & 49915 & 13933 & 0 &0.361 & 2.77  & 1597.1 &&1.11  &1110 & 521.7 \\ 
729&\{3,3,3,3,3,3\}       & 72808 & 22242 & 0 &0.465 & 2.15  & 1566.5 &&-&-&-  \\ 
768&\{2,2,3,2,2,2,2,2,2\} & 74822 & 18351 & 0 &0.518 & 1.93  & 1483.6 &&2.096 &1110 & 366.5\\
\hline
\end{tabular}
\label{tab:utilizationMK}
\end{table*}
The proposed scheme improves latency in the range of 63\% to 75\%. In terms of the operating frequency, our architecture operates in 2 to 3 orders of magnitude less than the architecture of \cite{Coppolino}. The lower operating frequency is directly translated to dynamic power saving. Finally, the proposed decoder offers $1.68\times$ to $3.05\times$ higher throughput concerning codes of \cite{Coppolino}.

The FPGA utilization and timing performance of various state-of-the-art decoders implemented under SC \cite{Coppolino}, unrolled SC \cite{Giard2016}, fast-SSC \cite{Rezaei2022} and MK fast-SSC \cite{Rezaei2022MK} is summarized in Table \ref{tab:utilization}. 
\begin{table*}
\centering
\caption{FPGA utilization and performance comparison of polar codes of size $1024$.}
\begin{tabular}{cccccccccccccc}
\hline
Work &  Algorithm &Rate& $P_e$ &IC Type&\begin{tabular}[c]{@{}c@{}}Tech.\\(nm)  \end{tabular}& \begin{tabular}[c]{@{}c@{}}$Q$\\$(Q_i, Q_c)$  \end{tabular} &LUTs & Reg. &  \begin{tabular}[c]{@{}c@{}}RAM\\(bits)  \end{tabular}  & \begin{tabular}[c]{@{}c@{}}L\\(MHz)  \end{tabular}&\begin{tabular}[c]{@{}c@{}}f\\($\mu$s)  \end{tabular}&\begin{tabular}[c]{@{}c@{}}T/P\\(Mbps)  \end{tabular}&\begin{tabular}[c]{@{}c@{}}\# of Supp.\\Codes  \end{tabular} \\ \hline
\cite{Coppolino}$^*$ & SC&$1/2$&60 & ASIC &65& (6, 6)&  -   &     -   & 29048   &  1.21 &1803.75  &   588.2&15\\
\cite{Giard2016} & Unrolled SC&$1/2$ & - & FPGA &40&  (5, 5) &   93225  &    43236   &  0 & 1.52&239 & 1500 & 15\\
\cite{Rezaei2022} & Fast-SSC&$1/2$ & 128  & FPGA &40& (7,6) &   18982  &     3384     & 37700  &  1.3&94.36 &  393& 15\\
\cite{Rezaei2022MK} & MK fast-SSC &$1/2$& 120  & FPGA &40& (5,4) &   23126  &     4548     & 40200  & 1.18& 86.14&  432& 55\\
this work & MK Comb. SC &any& - & FPGA &40& (5,5) &102740  &25139& 0  & 0.84& 1.19 &  1214.2 &55\\\hline
\end{tabular}\\
\flushleft 
\footnotesize{$~~~~~^*$ Similar to \cite{ercan2020practical}, the performance parameters are normalised to 40 nm CMOS technology based on scaling techniques from \cite{GiardTechMap}.}
\label{tab:utilization}
\end{table*}
It should be noted that only \cite{Rezaei2022MK} supports MK polar codes. To have a fair comparison, all designs are either implemented or scaled to 40 nm technology using the scaling techniques from \cite{GiardTechMap}.
Our scheme consumes an almost equal amount of memory as \cite{Coppolino} where it offers 31\% lower latency, 3 orders of magnitude lower operating frequency, and $2.07$ times the throughput. 
With respect to \cite{Giard2016}, both schemes occupy roughly the same amount of logic (LUTs) where \cite{Giard2016} consumes 72\% more registers. In terms of timing performance, \cite{Giard2016} has 81\% and 2 orders of magnitude higher latency and operating frequency, respectively. It however gains 23.5\% throughput. In comparison to designs based on fast-SSC \cite{Rezaei2022} and \cite{Rezaei2022MK}, the proposed scheme consumes approximately $5 \times$ logic and needs nearly 20 Kbits more registers. Note however that our decoder utilizes no RAM, while \cite{Rezaei2022} and \cite{Rezaei2022MK} require 36.8 and 39.26 Kbits of RAM, respectively. Furthermore, the proposed decoder achieves $2.09\times$ and $1.81\times$ higher throughput in comparison to \cite{Rezaei2022} and \cite{Rezaei2022MK}, respectively. Finally, our proposed decoder and \cite{Rezaei2022MK} support $55$ different block lengths, however, other decoders support only $15$ different codes. 
\section{{Conclusion}}
\label{sec_conc}
In this paper, we proposed a combinational-logic-based hardware architecture for decoding MK polar codes based on the SC algorithm. The proposed architecture offers a high throughput supporting an online rate assignment mechanism. It can decode an entire codeword in only one clock cycle which lowers the operating frequency and dynamic power consumption with reference to the synchronous SC-based architecture.
FPGA utilization for a variety of block lengths and kernel orderings is reported. Based on the implementation results, the proposed decoder can obtain the coded throughput of up to 1664.5 Mbps for a code of size $N=81$. A complexity analysis is also provided which verifies the implementation results. 

Finally, we built a Python-based polar compiler that can automatically generate 
the VHDL files needed for the FPGA implementation of the proposed decoders. By entering the block length and its kernel order, the polar compiler simply outputs all required VHDL modules automatically. The compile time for the polar compiler is also provided. 

\section*{Acknowledgment}
This research has been supported by the Academy of Finland, 6G Flagship program under Grant 346208. 


\begin{IEEEbiography}[{\includegraphics[width=1in,height=1.25in,clip,keepaspectratio]{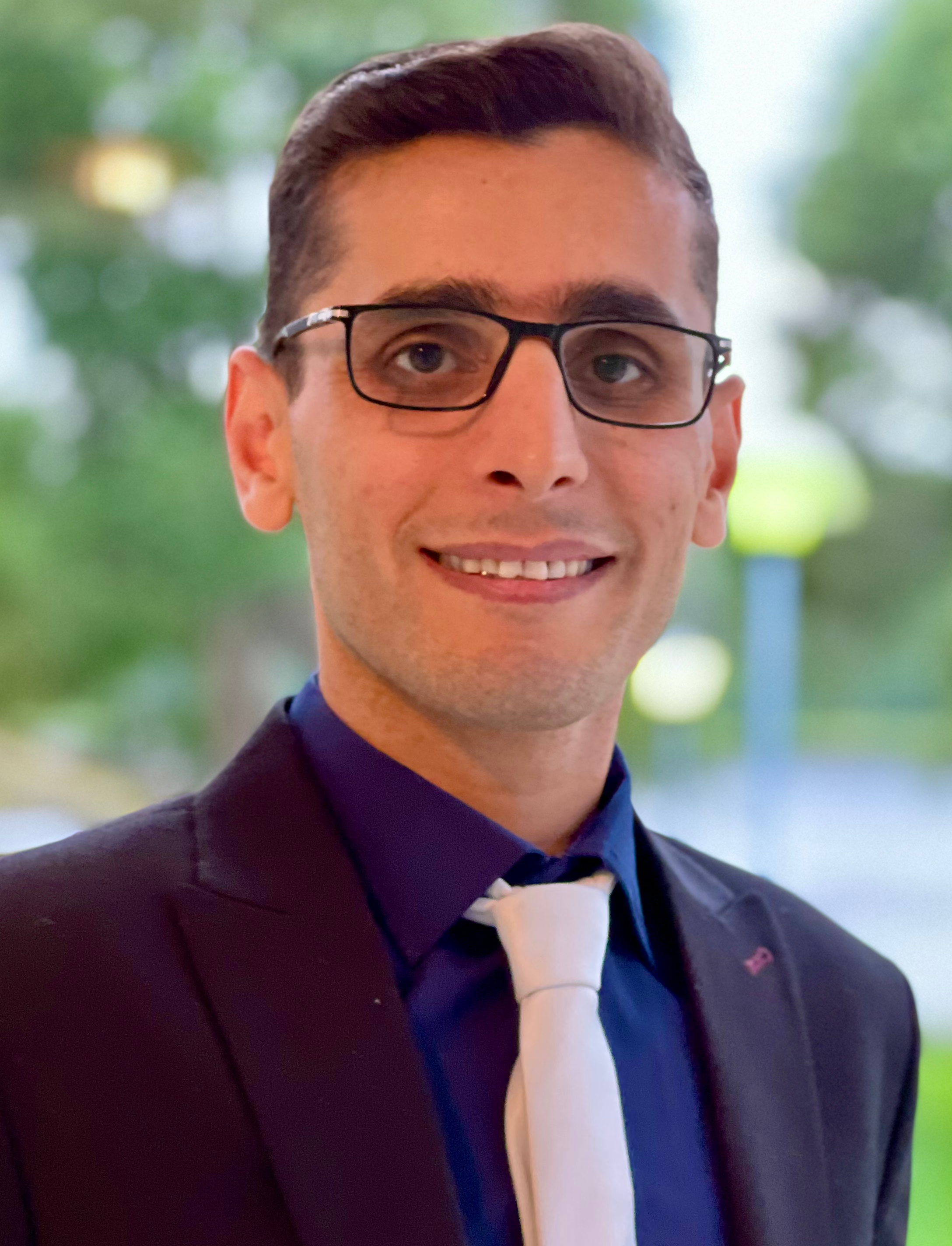}}]{Hossein Rezaei} received his M.Sc. degree in digital electronics from Iran University of Science and Technology, Tehran, Iran, in 2016. He is currently working toward the Ph.D at the University of Oulu, Oulu, Finland. His current research interests include design and implementation of error-correcting algorithms with a focus on Polar codes, VLSI design for digital signal processing, and implementation of communication systems on embedded platform.
\end{IEEEbiography}

\begin{IEEEbiography}[{\includegraphics[width=1in,height=1.25in,clip,keepaspectratio]{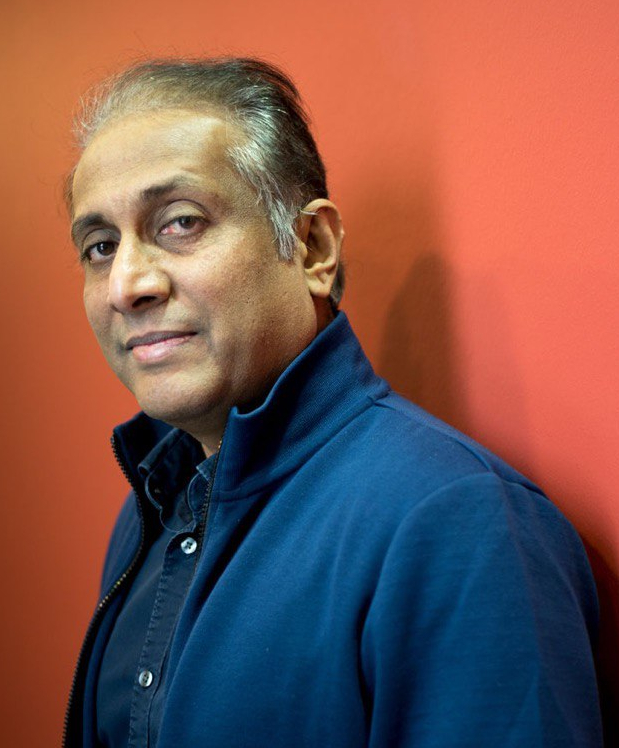}}]{Nandana Rajatheva}
(Senior Member, IEEE) received the B.Sc. (Hons.) degree in electronics and telecommunication engineering from the University of Moratuwa, Sri Lanka, in 1987, and the M.Sc. and Ph.D. degrees from the University of Manitoba, Winnipeg, MB, Canada, in 1991 and 1995, respectively. He is currently a Professor with the Centre for Wireless Communications, University of Oulu, Finland. During his graduate studies, he was a Canadian Commonwealth Scholar in Manitoba. From 1995 to 2010, he held a professor/associate professor positions with the University of Moratuwa and the Asian Institute of Technology, Thailand. He is currently leading the AI-driven Air Interface Design Task in Hexa-X EU Project. He has coauthored more than 200 referred articles published in journals and in conference proceedings. His research interests include physical layer in beyond 5G, machine learning for PHY and MAC, integrated sensing and communications as well as channel coding. 
\end{IEEEbiography}

\begin{IEEEbiography}[{\includegraphics[width=1in,height=1.25in,clip,keepaspectratio]{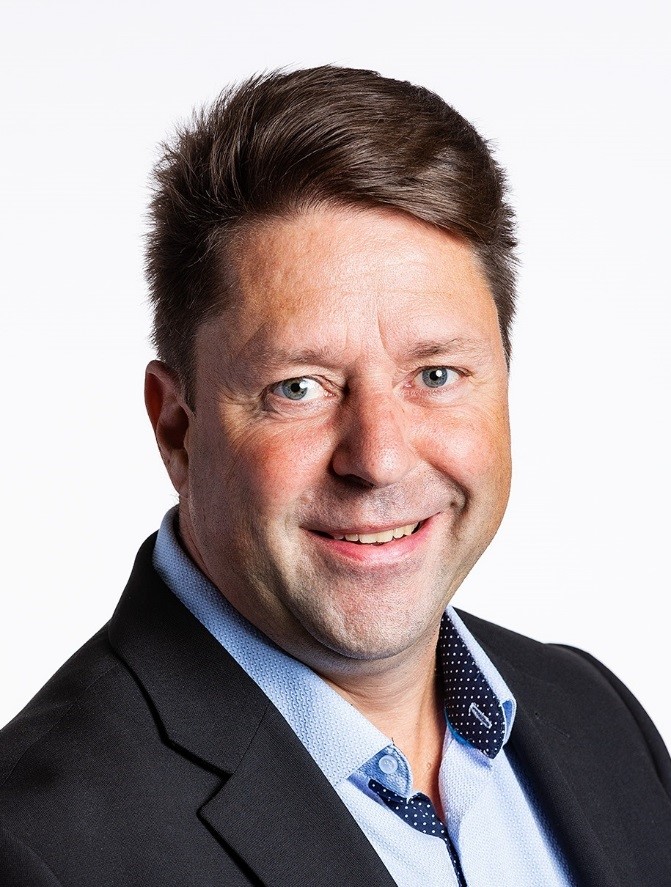}}]{Matti Latva-aho}
(Senior Member, IEEE) received the M.Sc., Lic.Tech. and Dr. Tech (Hons.) degrees in Electrical Engineering from the University of Oulu, Finland in 1992, 1996 and 1998, respectively. From 1992 to 1993, he was a Research Engineer at Nokia Mobile Phones, Oulu, Finland after which he joined Centre for Wireless Communications (CWC) at the University of Oulu. Prof. Latva-aho was Director of CWC during the years 1998-2006 and Head of Department for Communication Engineering until August 2014. Currently he serves as Academy of Finland Professor and is Director for National 6G Flagship Programme. He is also a Global Fellow with Tokyo University. His research interests are related to mobile broadband communication systems and currently his group focuses on 6G systems research. Prof. Latva-aho has published over 500 conference or journal papers in the field of wireless communications. He received Nokia Foundation Award in 2015 for his achievements in mobile communications research.
\end{IEEEbiography}
\vfill

\end{document}